\documentclass{raa}

\newcommand{\HI}{H\,{\sc {i}}~}
\newcommand{\kms}{km\,s$^{-1}$}

\newcommand{\Teff}{T$_{\rm eff}$}

\newcommand{\Msold}{M$_{\odot}$\,yr$^{-1}$}
\newcommand{\Mdot}{M$_{\odot}$\,yr$^{-1}$}

\newcommand{\Vlsr}{V$_{\rm lsr}$}
\newcommand{\Vexp}{V$_{\rm exp}$}
\newcommand{\Vcen}{V$_{\rm HI}$}
\newcommand{\Vsp}{V$_{\rm 3D}$}
\newcommand{\rin}{r$_{\rm in}$}
\newcommand{\rf}{r$_{\rm f}$}
\newcommand{\rout}{r$_{\rm out}$}
\usepackage{epsfig}
\usepackage{graphicx,times}             

\begin{document}

   \title{\HI emission from the red giant Y CVn with the VLA and FAST
}

   \volnopage{Vol.0 (200x) No.0, 000--000}      
   \setcounter{page}{1}          

   \author{D.~T. Hoai
      \inst{1,2}
   \and P.~T. Nhung
      \inst{1,2}
   \and L.~D. Matthews
      \inst{3}
   \and E. G\'erard
      \inst{4}
   \and T. Le~Bertre
      \inst{2}
   }

   \institute{Department of Astrophysics, Vietnam National Satellite
     Center, VAST, 18 Hoang Quoc Viet, Ha Noi, Vietnam
        \and 
             LERMA, CNRS, Observatoire de Paris/PSL, Sorbonne 
Universit\'es, F-75014 Paris, France; {\it thibaut.lebertre@obspm.fr}\\
        \and
             MIT Haystack Observatory, Off Route 40, Westford, MA
             01886, USA\\
        \and
             GEPI, CNRS, Observatoire de Paris/PSL, 
             92195 Meudon Cedex, France \\
   }

   \date{Received~~2009 month day; accepted~~2009~~month day}

\abstract{ Imaging studies with the VLA have revealed \HI emission associated 
with the extended circumstellar shells of red giants. We analyse 
the spectral map obtained on Y\,CVn, a J-type carbon star on the AGB. 
The \HI line profiles can be interpreted with a model of 
a detached shell resulting from the interaction of a stellar outflow with 
the local interstellar medium. We reproduce the spectral map by introducing 
a distortion along a direction corresponding to the star's motion in space. 
We then use this fitting to simulate observations expected from the FAST 
radiotelescope, and discuss its potential for improving our
description of the outer regions of circumstellar shells.
}

   \authorrunning{Hoai et al. }            
   \titlerunning{\HI from Y CVn }  

   \maketitle

%
%
\section{Introduction}           
\label{intro}

Asymptotic Giant Branch (AGB) stars are undergoing substantial mass
loss, and are often 
surrounded by extensive circumstellar envelopes. This phenomenon
dominates the last phases of the life of the majority of stars, and
is important for the evolution of the interstellar medium (ISM) and, 
more generally, for the chemical evolution of galaxies. 

The injection of stellar matter in the ISM occurs at large distances
from the central stars in regions where molecules (apart from possibly 
H$_2$, Morris \& Jura 1983) have not survived 
photodissociation by the UV photons of the interstellar radiation
field. Dust (Cox et al. 2012), H$_2$ (Martin et al. 2007), and atomic 
hydrogen (H\,{\sc {i}}, G\'erard \& Le\,Bertre 2006) are the main
tracers of these regions. Only the latter provides the high spectral 
resolution which is needed for describing the kinematics in these regions. 
However, the emission is weak and often confused by the competing
emission from the ISM on the same lines of sight. 

Several sources have been detected in \HI by the Nan\c cay Radiotelescope 
(NRT; G\'erard \& Le\,Bertre 2006), the Very Large Array (VLA; Matthews et
al. 2013), and the Green Bank Telescope (Matthews et al. 2015). 
In general the \HI emission is observed with a narrow line profile 
($\leq$\,5\,\kms), sometimes accompanied with a wider component 
(G\'erard \& Le\,Bertre 2006). 
The narrow line profiles provide direct evidence for the slowing down
of the stellar winds by their local ISM. 
The broad components should trace the central regions of the
outflows. Furthermore, the images 
often reveal a ``head-tail'' morphology (Matthews et al. 2008) 
indicating a distortion of the external shells 
resulting from the motion of the star through the ISM. 

The new generation of radiotelescopes that is under development is 
expected to transform this field of research, thanks to their high 
sensitivity and improved spatial resolution. In particular, 
the Five-hundred-meter Aperture Spherical radio Telescope (FAST) 
has a combination of angular resolution and sensitivity to large-scale
emission that is well-matched to the needs for studying \HI shells
around nearby ($\leq$\,1~kpc) AGB stars. 
To examine this potential we selected one source well characterized 
in \HI thanks to observations obtained with the NRT and the VLA. 
In particular the VLA data allow us to prepare simulations of what 
should be obtained with FAST. 

\section{Y CVn}
\label{sectgeneral}

Y CVn is a carbon star of the peculiar J-type, i.e. with a relatively
high $^{13}$C abundance ($^{12}$C/$^{13}$C  = 4; Abia \& Isern 1997), 
and no evidence of technetium (Little et al. 1987).
Basic parameters have been compiled in Table 1. For the distance, 
we adopt the parallax from Hipparcos (van Leeuwen 2007), 
and scale all other parameters to the corresponding distance.  
IRAS has revealed a detached shell (Young et al. 1993), that has been 
imaged at 90 $\mu$m by ISO (Izumiura et al. 1996). At a distance of 321 pc, 
it appears with an inner radius of 0.26 pc (2.8$'$), and outer one 
of 0.48 pc (5.1$'$). 
The CO line profiles indicate a wind with an expansion velocity $\sim$ 8 \kms. 
The mass loss rate in Table 1 is taken from Knapp et al. (1998)
and scaled to the adopted distance.

Le\,Bertre \& G\'erard (2004), observing with the NRT
(beamwidth: 4$'$ in right ascension and 22$'$ in declination), 
reported \HI emission at 21 cm in a line peaking at $\sim$ 20.5 \kms, 
close to the centroid of the CO line (cf. Table 1). The line profile 
is composite, with a narrow component of width $\sim$\,2.9\,\kms 
~(FWHM, Full Width at Half Maximum), and a broad component of width 14.3 \kms. 
The narrow component has been interpreted as an evidence of the slowing 
down of the wind revealed by the CO emission, and detected in \HI through the 
broad component. 

Libert et al. (2007) developed a spherical model in which the stellar 
outflow is abruptly slowed down from $\sim$ 8 \kms ~to $\sim$ 2 \kms ~at a 
termination (or reverse) shock which 
is located at the inner radius (\rin)
of the detached shell revealed by IR images. 
The forward shock, where external matter is compressed by 
the expanding shell, is located at the outer radius (\rout) 
of the IR detached shell. 
In this picture, a detached shell is formed by compressed
stellar and interstellar material separated by an interface 
located at \rf,  
where the two media are in contact (Lamers \& Cassinelli 1999). 
This approach allowed Libert et al. to fit the \HI line profiles obtained 
at various positions with the NRT. Some technical details are given 
in the Appendix.

Y CVn has been observed in \HI with the VLA in the D configuration 
(i.e. with baselines up to 1.0\,km) by Matthews et al. (2013, M2013). 
They present a map with a much better spatial resolution ($\sim 1'$) 
than that offered by the NRT. The \HI shell appears offset with
respect to the star 
position by $\sim$\,1$'$ in a direction opposite to the stellar
motion. It shows a broken ring structure with a dearth of emission
ahead of the star. The spatially integrated emission globally coincides 
with the IR emission at 90\,$\mu$m detected by ISO, even though it differs 
in some details (M2013), in particular with a less circular and more clumpy
\HI emission.

Although the integrated line profile agrees with that used by Libert 
et al. (2007) the original model developed by Libert et al. does 
not agree well with the spatial distribution of the \HI emission 
observed at the VLA (M2013). However using
the same modeling approach, but adjusting the mass loss rate and 
the duration, Hoai (2015) could obtain a better fit of the circularly
symmetrized \HI emission. 

\begin{table}
{\centering
\caption{Properties of the central star}
\begin{tabular}{lll}
\hline
                     & Y CVn       & reference \\
\hline
spectral type        & C5,4J(N3)   &  General Catalogue of Variable Stars (Samus+ 2007-2013) \\
pulsation period (d) & 157         &  idem \\
\Teff (K)            & 2760        &  Bergeat et al. (2001) \\
d (pc)               &  321        &  Hipparcos (van Leeuwen 2007) \\
\Vlsr (\kms)         &   21.1      &  Knapp et al. (1998) \\
\Vexp (\kms)         &    7.8      &  idem \\
\.M (\Mdot)          &    2.4$\times10^{-7}$     &  idem \\
\Vsp (\kms), PA(deg) & 35.4, 30.1  &  Matthews et al. (2013) \\
\Vcen (\kms)         &  20.6       &  Le Bertre \& G\'erard (2004)\\
\hline
\end{tabular}\\}
\label{basicdata}
\end{table}

\section{\HI modeling of the VLA spectral map}
\label{sectdetailed}

\begin{figure}
\centering
   \includegraphics[width=15.8cm, angle=0]{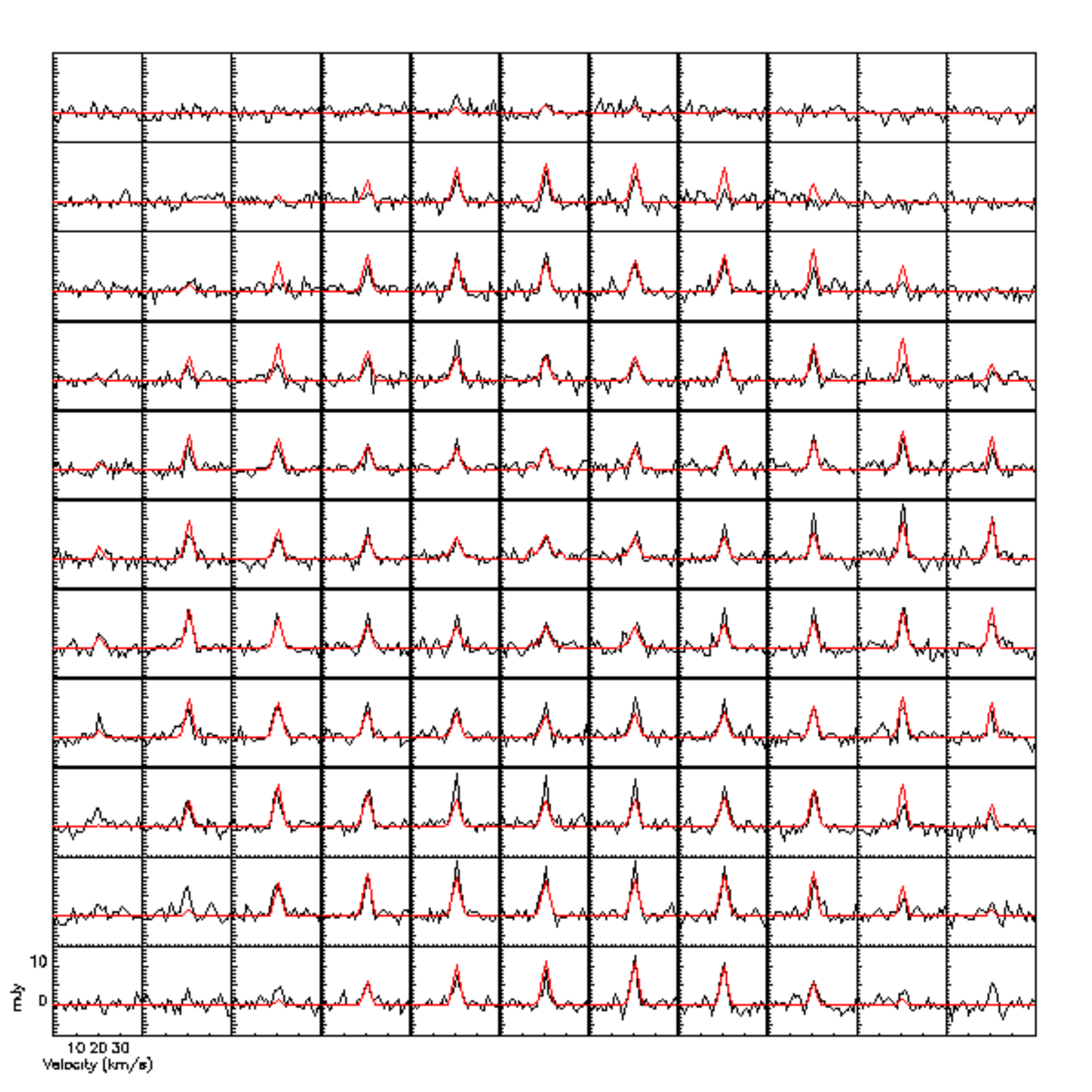}
\caption{Y CVn: comparison between the spectral map observed at the VLA 
(black lines) and the synthesized spectral map (red lines). The steps
are 50$''$ in right ascension and declination. The grid is centered
on the stellar position.}
  \label{spectralmapYCVn}
\end{figure}

A spectral map based on the observations of M2013 
was derived by extracting spectra over a grid of
14$'\times$14$'$, centered roughly on the star with steps of 50$''$ 
in right ascension and declination, corresponding to approximately 
one beam diameter. The channel width corresponds to 1.28\,\kms. The 
spectral map is shown in Fig.~\ref{spectralmapYCVn} (black lines).

We start from the spherical model developed by Libert et al. (2007),
and improved by Hoai et al. (2015). The stellar effective temperature 
is larger than 2500\,K (cf. Table 1), 
implying that hydrogen is mainly in atomic form in the atmosphere 
and outwards (Glassgold \& Huggins 1983).
We adopt a constant mass loss rate, with a value as close as possible to 
that obtained from CO estimates (Table 1), and assume that all
hydrogen is in atomic form. 
For the expansion velocity of the free wind we adopt the value obtained 
from the CO line profile. An arbitrary temperature profile
is adopted between the two limits, \rin ~and \rout, with a logarithmic 
dependence from \rin ~to \rf, and a constant temperature from \rf 
~to \rout. 

As explained in Sect.~\ref{sectgeneral}, the circularly symmetrized
\HI emission can be reproduced with the Libert et al. model, subject
to a reduction of the mass loss rate (by a factor 2 relative 
to the estimate from CO data) and a corresponding increase of 
the duration for the detached shell formation (Table~\ref{parameter}, 
Hoai 2015). An average mass loss rate lower than the present day
value may mean that the hypothesis of a constant mass loss rate over
$\sim 10^6$\,years is too simplistic. (However, the change that is now
implied is much less than the one of about two orders of magnitude, 
initially suggested by Izumiura et al. (1996).)

\begin{table}
\centering
\caption{Parameters used for the \HI modeling (d\,=\,321\,pc)}
\begin{tabular}{ll}
\hline
\.M (\Mdot)         &   1.3$\times10^{-7}$\\
duration (yrs)      &   7$\times10^5$\\
\Vexp ~(\kms)       &   8.0      \\
\rin ~(arcmin, pc)  &   2.8, 0.26\\
\rf ~(arcmin, pc)   &   4.0, 0.37\\
\rout ~(arcmin, pc) &   5.1, 0.48\\
temperature index   &  --6.0     \\
\hline
\end{tabular}\\
\label{parameter}
\end{table}

\begin{figure}
\centering
\epsfig{figure=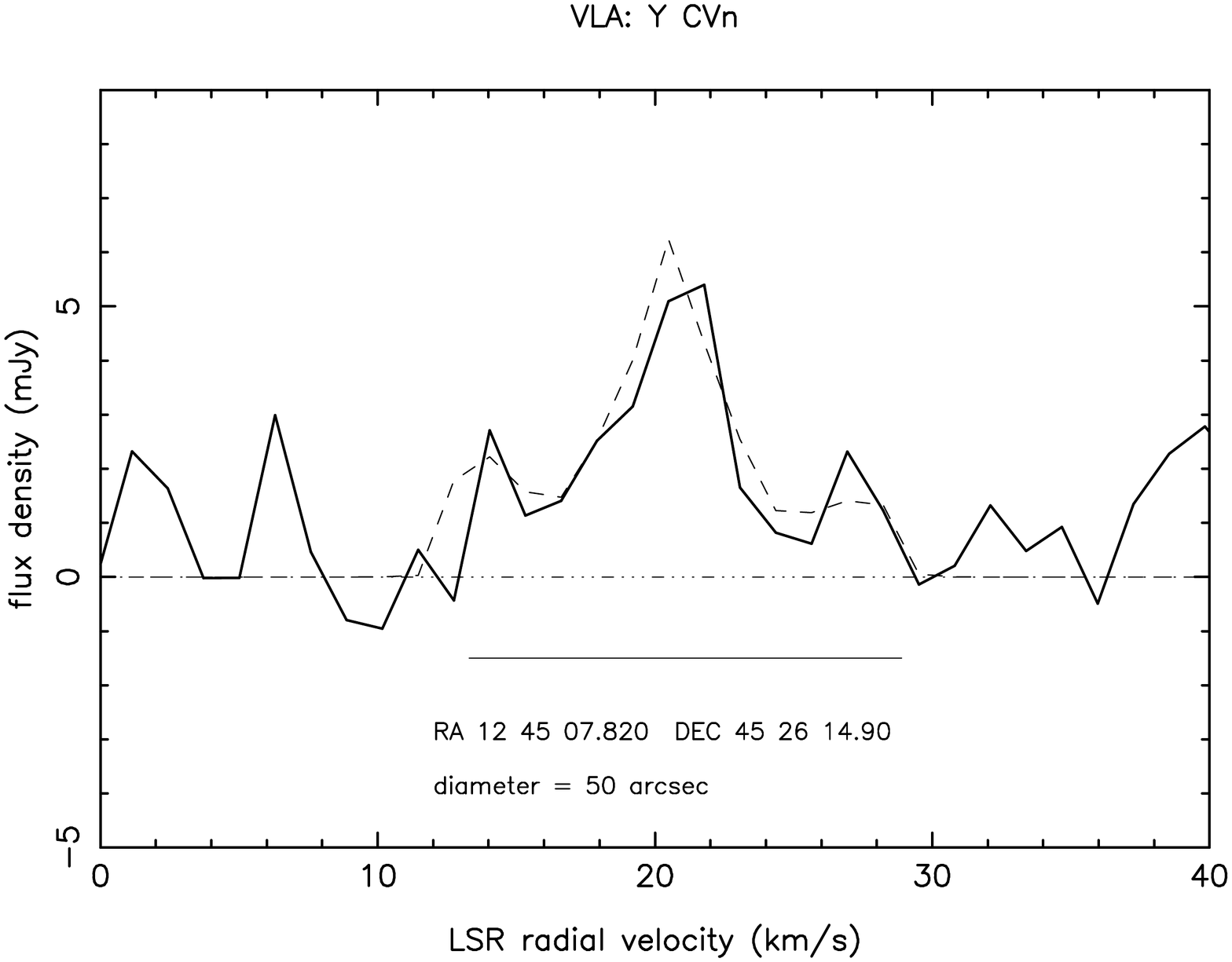,angle=0,width=8cm}
\caption{Spectrum obtained by the VLA on the stellar 
position (enlargement of the central spectrum in 
Fig.~\ref{spectralmapYCVn}). The horizontal bar represents the
velocity range covered by the CO emission (21.1$\pm$7.8\,\kms), and 
the dashed line the fit discussed in Sect.~\ref{sectdetailed}.}
  \label{VLAcentralpanel}
\end{figure}

In order to reproduce the distortion in the direction of motion 
observed in the VLA image of Y\,CVn, we also apply a geometrical 
factor to the distance of each parcel of gas from the central star, 
such that the morphology becomes elongated along this direction. 
For that purpose, we apply a factor 
(1\,+\,a\,sin\,$\phi$), with $\phi$ being counted from a plane 
perpendicular to the direction of motion and towards this direction.  
A good fit of the spectral map (red lines in Fig.~\ref{spectralmapYCVn}), 
obtained by a least-square minimization of all spectra 
simultaneously, gives a\,=\,$-$0.17. The general trends are reproduced, 
although not the details such as the broken structure of the ring.
The narrow spectral component dominates the map with a peak flux 
density of $\sim$\,10\,mJy. 
In addition, in the central panel of the map, one can identify two
side peaks of $\sim$\,3\,mJy at 14 and 28\,\kms ~(see also 
Fig.~\ref{VLAcentralpanel}), separated by $\sim$\,14\,\kms 
~($\approx$\,2$\times$\Vexp). Although at the limit of detection 
($\sigma \sim$\,1.3\,mJy), they could reveal the freely flowing stellar wind. 
As the line profile is not rectangular, it suggests that 
the corresponding region is spatially resolved. 
The size would thus be at least $\sim1'$, already much larger 
than the size of the region traced by CO ($\sim13''$, Neri et al. 1998).

\section{Simulations for FAST}

\begin{figure}
\centering
\epsfig{figure=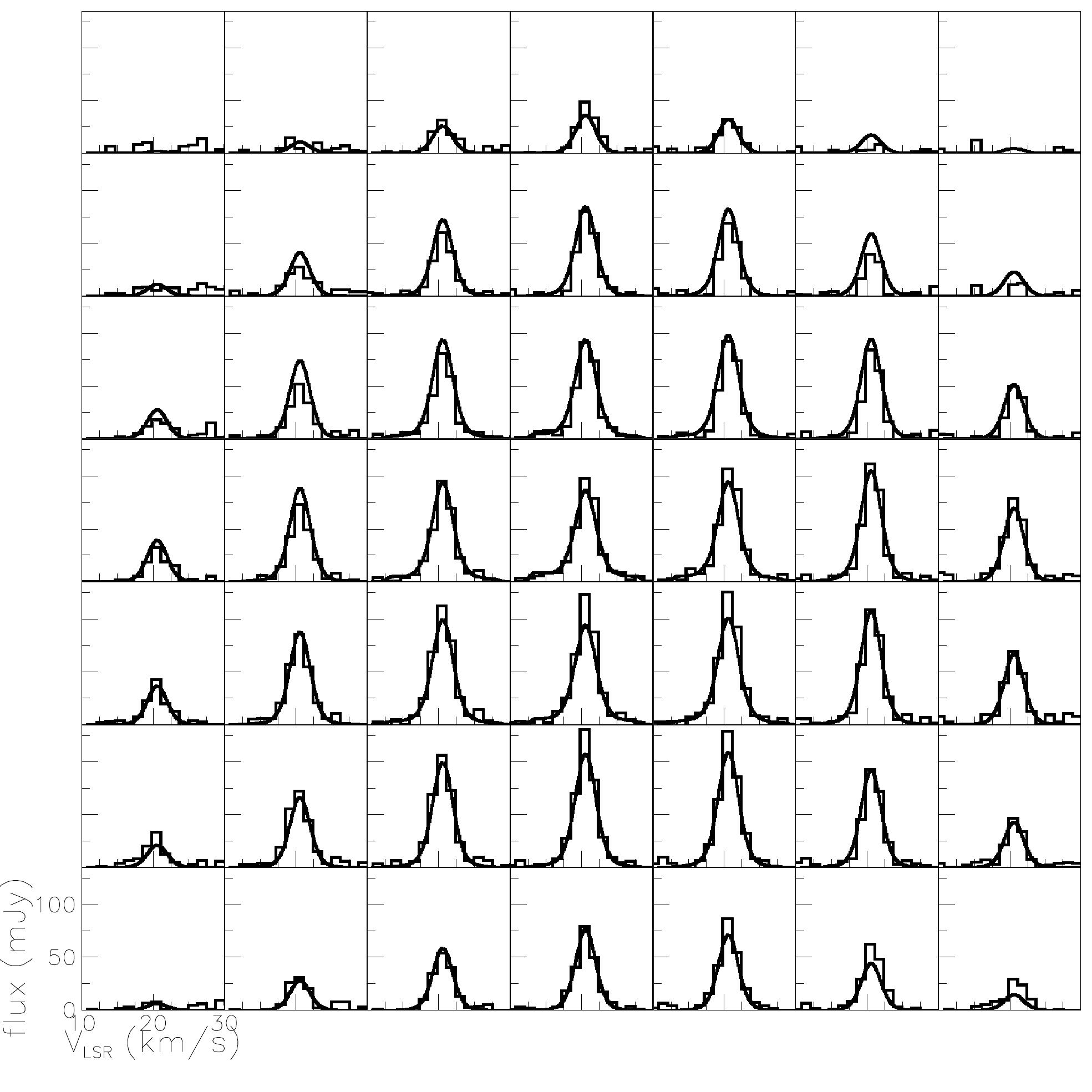,angle=0,width=8cm}
\caption{Y CVn: simulation of the spectral map observed with FAST. The steps
are 1.5$'$ in right ascension and declination. The grid is centered
on the stellar position. For comparison, the data from the VLA have
been convolved by a gaussian beam of 2.9$'$ and are displayed as 
histograms.}
  \label{simulFAST}
\end{figure}

FAST has an illuminated aperture of 300\,m (Nan et al. 2011). 
To simulate an observation of Y CVn, we convolve the model obtained 
in Sect.~\ref{sectdetailed} by the response of a FAST beam at 21\,cm. 
We adopt a gaussian beam profile of FWHM\,=\,2.9$'$.
The result is shown in Fig.~\ref{simulFAST}.

Observing Y CVn with FAST will allow us to determine the line profiles
in much better detail than with the VLA. This is important for
constraining the kinematics and the physical properties of the gas 
in the detached shell. It would be interesting also to search for
emission outside the limits defined by the image obtained at 90\,$\mu$m
by ISO: M2013 noted that, although similar in the main features, the \HI 
and the ISO images differ in the details. A spectral characterization of the
differences will allow us to understand better the coupling between 
gas and dust.  

The high sensitivity reached by FAST would allow us to investigate the
presence of gas emitting outside the velocity range defined by CO
observations. In the case of Mira, Matthews et al. (2008) report a
gradient of velocity along the tail with a centroid velocity trailing from 
+45\,\kms ~on the stellar position, to +22\,\kms ~2\,degrees
away. Unpublished spectra from the NRT on Y CVn suggest the presence of
an \HI emission between +6\,\kms ~and +12\,\kms, possibly extending
south. 

\begin{figure}
\centering
\epsfig{figure=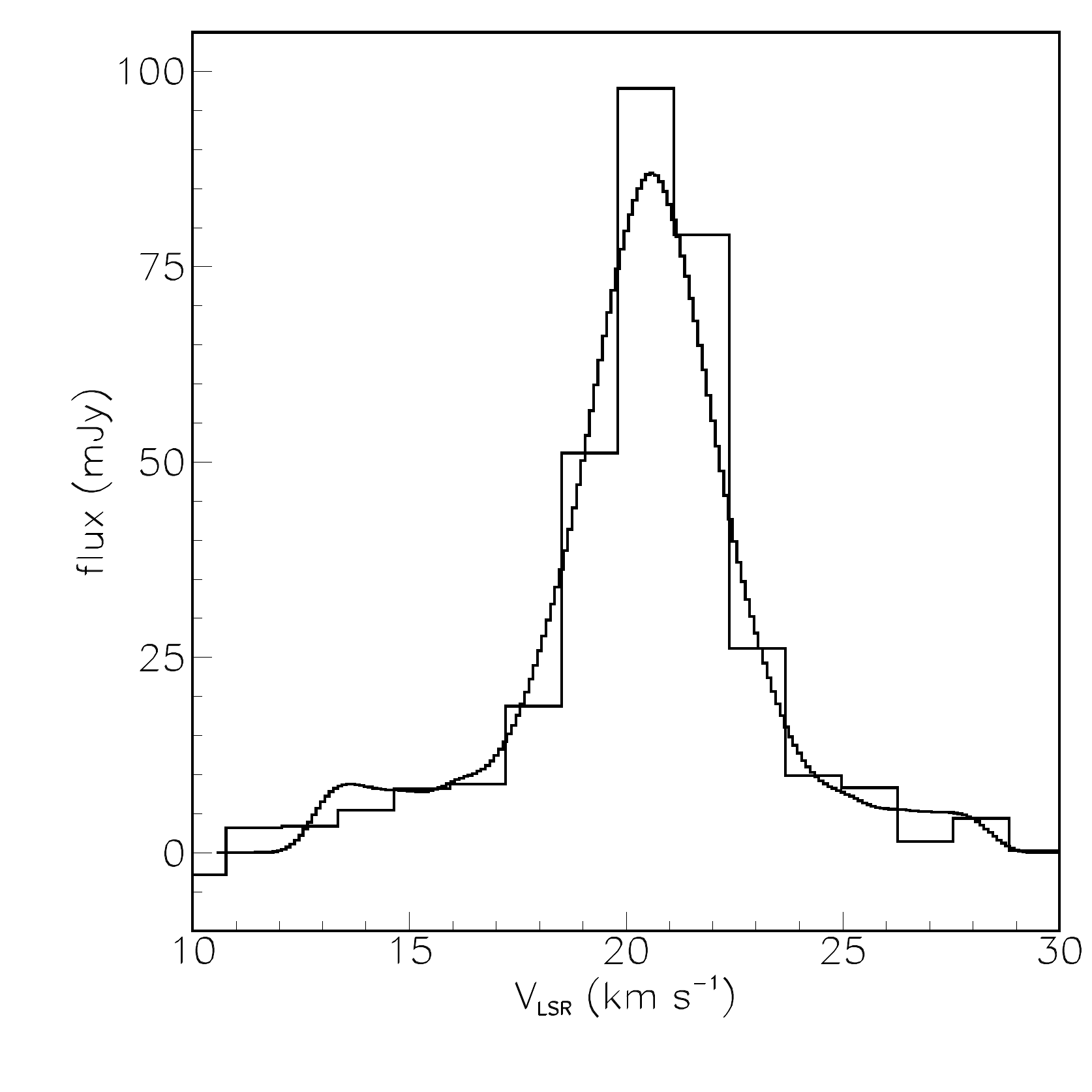,angle=0,width=7cm}
\caption{Simulation of the spectrum obtained by FAST on the stellar 
position (enlargement of the central spectrum in Fig.~\ref{simulFAST}). }
  \label{simulFASTcentralpanel}
\end{figure}

Finally the freely expanding wind will be better 
detected and in particular it should be possible to constrain 
the extent of its region, which presently is only assumed to be identical 
to the inner dust shell. 
The FAST beam width being comparable in size to the inner radius 
of the dust shell, a coupling of the model with \HI observations
should allow us to compare the relative positions of gas and dust. 
From the line profile, for which we predict 
asymmetrical wings (Fig.~\ref{simulFASTcentralpanel}), 
we can see that it is possible to also constrain the morphology of this 
region along the line of sight if a sensitivity of $\sim$\,2\,mJy 
over 5\,\kms ~could be reached. 
The broad component is  detected by the VLA on the line of sight to
the central star (Fig.~\ref{VLAcentralpanel}), and also when the spectrum is 
integrated in a circumstellar aperture of 2.8$'$ radius centered on the
star (M2013, Fig.~16).  
However, the VLA data have an insufficient
Signal-to-Noise Ratio for characterizing the details of the free
flowing wind region. On the other hand, the large collecting area of 
FAST can bring the high sensitivity which is
needed. 

\begin{figure}
\epsfig{figure=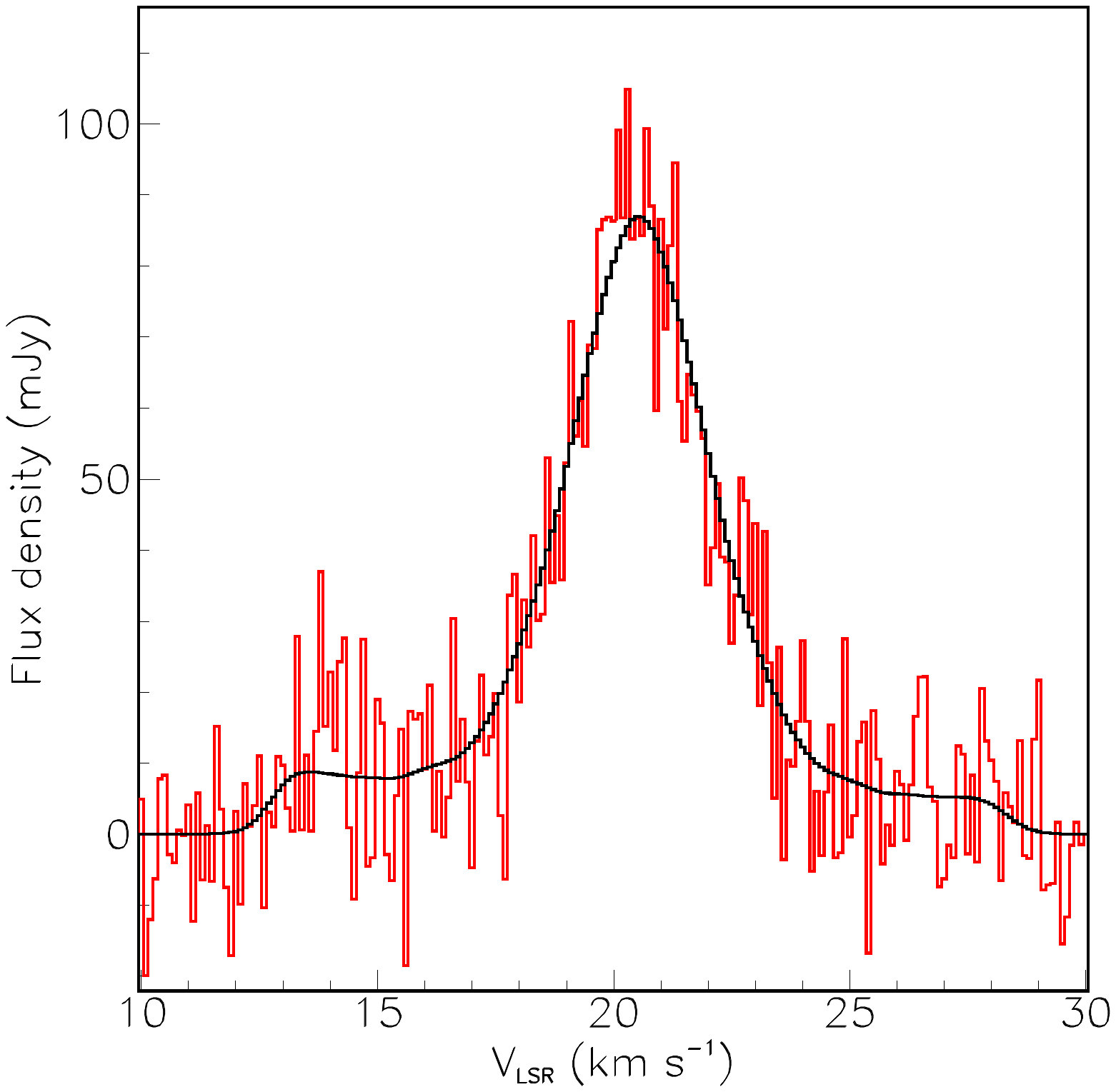,angle=0,width=8.5cm}
\epsfig{figure=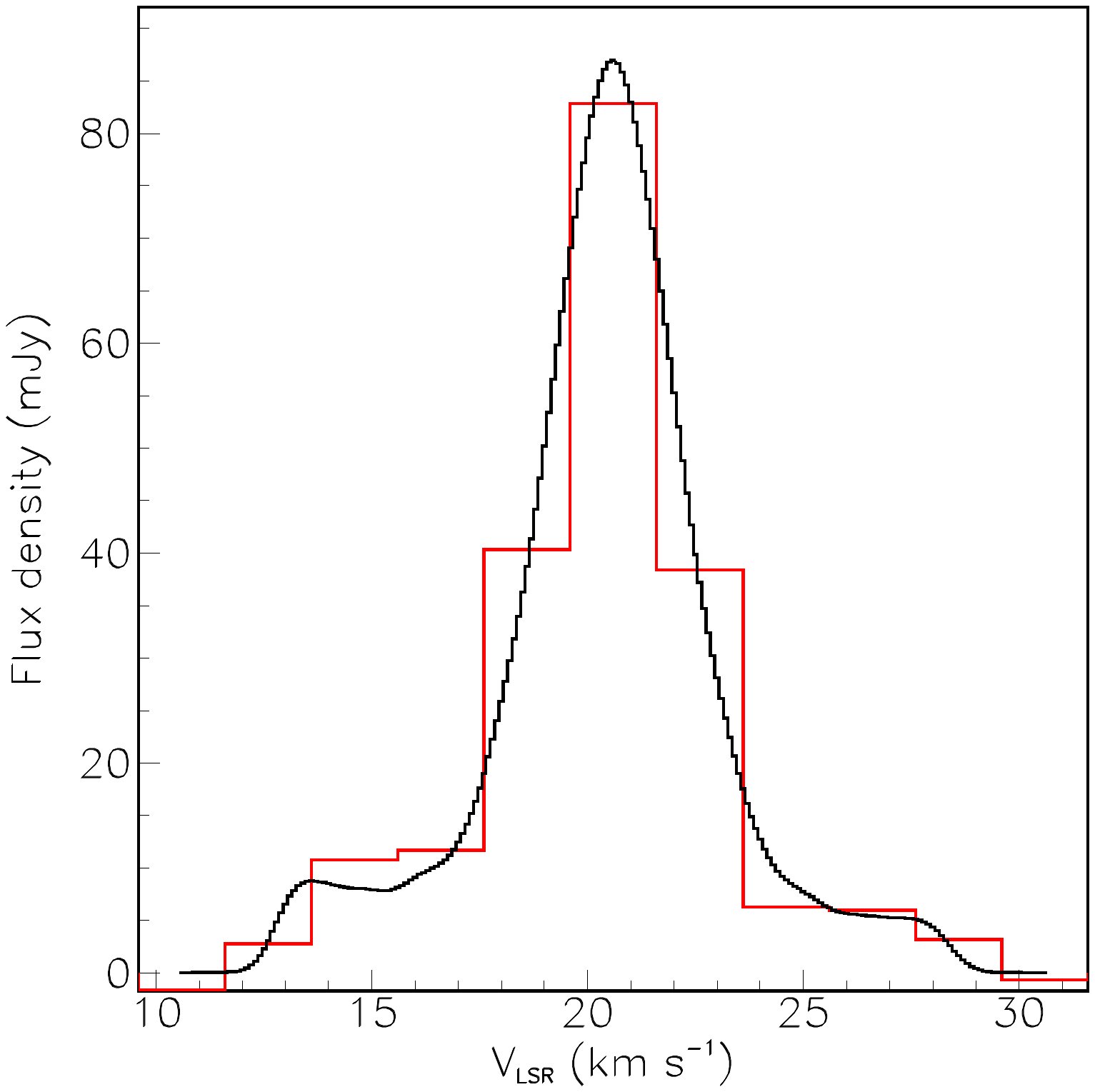,angle=0,width=8.5cm}
\caption{Same as in Fig.~\ref{simulFASTcentralpanel}, assuming a noise 
corresponding to 10\,mJy (1\,$\sigma$) per 0.1\,\kms ~channel (red
curve in left panel), 
and 2\,mJy per 2\,\kms ~channel (right panel).}
  \label{simulFASTcentralpanelplusnoise}
\end{figure}

Assuming a gain of 16\,K/Jy and a system temperature of 30\,K, a
sensitivity of 1\,$\sigma$\,=\,10\,mJy per 0.1\,\kms ~channel should 
be reached in 5 minutes (Fig.~\ref{simulFASTcentralpanelplusnoise}, 
left panel). This spectral resolution and sensitivity are 
needed to explore the properties of the detached shell. For the
free-flowing region a lower spectral resolution could be allowed. The
same sensitivity would translate to $\sim$\,2\,mJy over 2\,\kms,
enough to detect an asymmetry between the two wings of the broad
component (Fig.~\ref{simulFASTcentralpanelplusnoise}, right panel). 
A\,12$'\times12'$ map of Y\,CVn, with a 1.5$'$
step, corresponding to Nyquist sampling, and assuming a single receiver, 
would require 5-6 hours. This is only a rough estimate as the observing 
time would depend on the availability of a multi-beam receiver, the 
strategy of observation, and the exact sampling used in the
observations. In particular, Mangum et al. (2007) recommend a sampling
of at least twice Nyquist for on-the-fly imaging (i.e. 5 points
per FWHM rather than 2). The details of the observing programme should
be defined in concertation with FAST experts.

Being a giant single-dish radiotelescope, FAST will be optimized  
in surface brightness sensitivity for spatial structures filling its beam. 
It will thus be particularly suited for revealing the inner regions 
of nearby ($\leq$\,1\,kpc) circumstellar
shells which are traced by the broad components of the \HI emission
and whose sizes are expected to be a few arcminutes. 
Evolved stars with large mass loss rate (up to 10$^{-4}$\,\Msold) may
be surrounded by circumstellar shells with fast winds sweeping up the
ISM at large distances (up to 2.5 pc, Villaver et al. 2002). 
The \HI emission from 
these shells has been modeled by Hoai et al. (2015), but presently  
has not been detected unambiguously, possibly for an insufficient  
surface brightness sensitivity. 

\section{Prospects}

Our approach is empiric, but allows us to approximately reproduce 
the original VLA data. It thus can be useful for preparing simulations
of observations that could be obtained with the new generation of
radiotelescopes. Such simulations are useful because only a few observations 
with good signal-to-noise ratio are available. The radiotelescopes
presently under development will bring spatially resolved line
profiles of better quality for circumstellar shells around evolved
stars, and a phenomenological description can be useful for exploiting
the new data. However, it would be more satisfactory to have a
genuine physical modeling of these sources, not only for preparing 
the new observations but also for interpreting them. 

An important improvement would consist in the inclusion of a cooling law 
for determining the temperature profile of matter having crossed the 
termination shock instead of using an arbitrary temperature profile. 
It is important because for a subsonic wind, the
kinematics and the temperature distribution are coupled, and because 
in these objects the separation between the forward shock and the
reverse shock is comparable to the size of the system.

Another improvement would consist in modeling the effect of the motion 
of the star through the interstellar medium on the circumstellar shell
instead of using an arbitrary distortion of the circumstellar shell. 
Villaver et al. (2003, 2012) have developed such models. However, 
as shown by Hoai et al. (2015) the temperature of the gas in these models 
is too high to account for the narrow width of the observed line
profiles (which brings us back to the previous argument). Finally, 
stellar evolution incorporating mass loss rate and expansion velocity 
may also need to be taken into account (Villaver et al. 2002, 2012). 
Conversely, \HI observations should help to constrain it. 

\section{Conclusions}

We have developed simulations that allow us to predict \HI fluxes and line
profiles expected on a prototypical source that could be observed with FAST. 
The high sensitivity to \HI surface brightness, which is 
expected from this new facility, is opening up exciting prospects 
for the observations of extended shells around nearby  
red giants. 

An observation such as that proposed on Y CVn 
would provide an excellent illustration of the FAST
potential, and as well bring useful informations for constraining wind-ISM
interaction models.  The high spectral resolution should 
reveal kinematic effects presently not accessible in the VLA data. 
The high sensitivity should also give full access to the freely 
expanding wind region (which appears truncated by photodissociation 
when observed with molecular tracers) and allow us to constrain 
its morphology. 

Our approach can as well be used for making simulations 
of images that will be obtained with future interferometers such as 
the ngVLA (next generation VLA) and the SKA (Square Kilometre Array). 
However, it is somewhat limited by the
hypotheses that are presently made 
(arbitrary temperature profile and distortions of the circumstellar 
shells). Therefore, the development of hydrodynamic models would be 
useful, and as well, considerably improve our understanding of 
the interaction between stellar winds and the ISM.

\begin{acknowledgements}
We are grateful to P. Darriulat and P.\,N. Diep for their continuous
encouragements and stimulating comments. 
The VLA observations discussed here were part of NRAO program AM1001. 
Financial support is acknowledged from the Vietnam National Satellite 
Centre (VNSC/VAST), the NAFOSTED funding agency, the World Laboratory, 
the Odon Vallet Foundation and the Rencontres du Viet Nam. 
We also thank the PCMI programme of the CNRS for financial support.
\end{acknowledgements}

\section{Appendix}

A star undergoing mass loss gets surrounded by an expanding
circumstellar shell. However at some distance, the supersonic wind
interacts with external matter (Lamers \& Cassinelli 1999). It
produces a shell of denser material which has been detected around
many evolved stars through thermal emission by dust (Young et
al. 1993). For a long time the gas in these detached shells remained 
undetected, mainly because of the photodissociation of molecules by UV
photons from the interstellar radiation field. Following the detection
of atomic hydrogen in the detached shell of Y CVn (Le Bertre \&
G\'erard 2004) and other stars (G\'erard \& Le Bertre 2006), Libert et al. 
(2007) developed a simple model of the detached shell around Y CVn.

\begin{figure}
\centering
\epsfig{figure=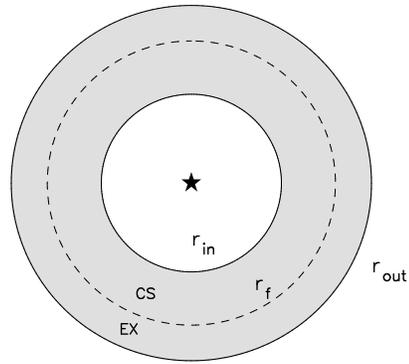,angle=0,width=10cm}
\caption{Sketch of an \HI detached shell. The termination shock is
 located in \rin, the contact discontinuity in \rf, and the forward shock
 in \rout. CS stands for circumstellar material, and EX for interstellar
 matter incorporated in the detached shell (adapted from Libert et al. 2007).}
  \label{detachedshell_schema}
\end{figure}

Libert et al. assume spherical symmetry, stationarity, and gas in atomic 
form,with 10 percent of $^4$He and 90 percent of H. A detached shell is
produced by the abrupt slowing down of the stellar flow at a termination 
shock (\rin, Fig.~\ref{detachedshell_schema}). At this inner boundary, 
the gas is heated to a temperature $\approx$ \Vexp$^2$ 
(Dyson \& Williams 1997). Beyond \rin, the gas is expanding at 
$\sim$\,1/4$\times$\Vexp, and then slower and slower as it cools down. 
Libert et al. (2007) adopt a logarithmic temperature profile. 
This assumption allows them to write in a simple way the equation of motion, 
which in this condition depends only on the logarithmic temperature index. 
Its value is selected such as to obtain a good fit to the 
\HI line profiles. The outer limit (\rout) of the detached shell 
is defined by the leading shock at which external matter is compressed
by the expanding shell. Between \rin ~and \rout
~the detached shell is composed of compressed circumstellar matter (CS
in  Fig.~\ref{detachedshell_schema}) and interstellar matter (EX
in  Fig.~\ref{detachedshell_schema}) separated by a contact
discontinuity located in \rf. The interstellar matter which is
incorporated in the detached shell corresponds to that originally
occupying the sphere of radius \rout ~at the galactic location of the
central star. The duration of the mass loss process is selected such as to
fit the spatially integrated intensity of the \HI emission, assuming 
a constant mass loss rate from the central star. The equation of
motion is solved numerically between the two limits of the detached
shell. Temperature, velocity and density are then used to calculate the
line profiles for any desired position on the sky and beam profile. 
Although crude, this modelling allowed Libert et al. (2007) to
reproduce the spatial distribution and line profiles of the \HI 
emission observed at low angular resolution by the NRT, adopting for
the mass loss rate of Y CVn a value obtained from CO observations
(``present day'' mass loss rate). 

However, the data obtained with a better spatial resolution by the VLA
(Matthews et al. 2013) did not agree as well, and led Hoai (2015) to 
relax the mass loss rate value, and correspondingly the age of the
detached shell. The new value is obtained through a least-square
minimization on the VLA spectral map. In the present work, we improve 
the quality of the fit by introducing an artificial deformation of the
detached shell shape (Sect.~\ref{sectdetailed}). 

\label{lastpage}

\end{document}